
\magnification=1200

\vsize = 24truecm
\hsize = 16truecm

\null
\vskip-1truecm
\rightline{IC/95/53}
\rightline{INTERNAL REPORT}
\rightline{(Limited Distribution)}
\vskip1truecm
\centerline{International Atomic Energy Agency}
\centerline{and}
\centerline{United Nations Educational Scientific and Cultural Organization}
\medskip
\centerline{INTERNATIONAL CENTRE FOR THEORETICAL PHYSICS}
\vskip2truecm
\centerline{{\bf INTEGRABLE CHERN-SIMONS GAUGE FIELD THEORY}}
\centerline  {{\bf IN 2+1 DIMENSIONS}}
\vskip1.5cm
\centerline{O.K.Pashaev\footnote{$^*$}{Permanent address: Joint Institute for
Nuclear Research,  Dubna 141980, Russian Federation. E-mail:
pashaev@main1.jinr.dubna.su}}
\centerline{ International Centre for Theoretical Physics, Trieste, Italy.}
\vskip1cm
\centerline{ABSTRACT}
\baselineskip=16pt
\medskip

The classical spin model in planar condensed media is represented as the $U(1)$
Chern-Simons gauge field theory. When the vorticity of the continuous flow
of the media
coincides with the statistical
magnetic field, which is necessary for the model's integrability, the
theory admits zero
curvature connection. This allows me to formulate the model in terms of
gauge - invariant
fields whose evolution is described by the Davey-Stewartson (DS) equations. The
Self-dual Chern-Simons solitons described by the Liouville equation are
subjected to  corresponding integrable dynamics. As a by-product the
2+1-dimensional zero-curvature  representation for the DS equation is obtained
as well as the new reduction conditions related to the DS-I case. Some possible
applications for the statistical  transmutation in the anyon
superfluid and TQFT are briefly discussed.
\vskip1cm
\centerline{MIRAMARE -- TRIESTE}
\centerline{April 1995}

\vfill\eject

\noindent
{\bf 0 Introduction}
\par
\medskip
Recently, the Chern-Simons (CS) solitons are derived as solutions of the
self-dual CS
system [1]. These solutions have an important quantum meaning producing
excitations
known as anyons possessing arbitrary statistics [2]. They are relevant to
the Quantum
Hall Effect
(QHE) and the
High Temperature Superconductivity
since they are still being intensively studied [3,4]. In fact, the
effective Ginzburg-Landau
theory of the QHE turns out to be described by the Nonlinear Schr\"odinger
Equation
(NLSE) in 2+1 dimensions for a complex matter field which couples with the
Abelian
Chern-Simons gauge
field. When the field's self-interaction coupling strength is
related to the Chern-Simons coupling constant the model reduces to the static
self-dual CS
equations [1]. These equations, being conformal invariant, are completely
integrable and
admit the Lax pair representation [5]. The time evolution of the model, in
general, can be
arranged as a conformal mapping of the plane to itself at different times. \par
But analysis of the time dependent solutions with integrable dynamics
remains an open problem. Jackiw
and Pi [6] mentioned the
tantalizing similarity of the problem with the Davey-Stewartson equation,
being a well
known integrable extension of the NLSE in 2+1 dimensions
[13]. \par
In the present paper I reduce the self-dual CS model from an integrable
model in 2+1
dimensions. This model was first derived by Ishimori and has integrable
dynamics of the
magnetic vortices [7]. Its formulation in terms of the vorticity for
continuous flow tends to
clarify the physical meaning of the model and allows an extension for
higher dimensions,
admitting the bilinear Hirota representation
[8,10]. The physical applications of the model are connected with the
resolution of the
anomalous behavior of the linear momentum in ferromagnets [9]. As it was
shown in the
latter paper by using the de-localized
electron model of ferromagnets, the canonical momentum becomes well defined
due to a
fermionic background.
\par The model considered in the present paper is a modification of the
classical
Heisenberg model for condensed media, having hydrodynamical flow with its
vorticity
related to a topological charge density. In this case a hydrodynamical vortex
in the
magnetic (anyon) fluid is also inducing a magnetic vorticity.
\par The CS gauge field theory is then obtained by projecting spin
variables on
the tangent plane to the sphere (spin phase space). This idea has been
recently applied
to the classical Heisenberg model in [12]. Furthermore, the stationary
magnetic vortices
of the model have been related to the Chern-Simons solitons, while the
topological
charge
of the former has been related to the electric charge of the latter. \par
In the case of equality between hydrodynamical vorticity and the
``statistical''
magnetic  field, the theory can be formulated for a zero curvature effective
gauge field. This allows  me to reformulate the model in terms of gauge
invariant matter fields. As shown, their  evolution is described by the DS
equations and at the same time the fields satisfy the CS system. For the
self-dual case, we reproduce CS solitons with an  integrable evolution
replacing the statistical gauge field with a velocity field. \par
As a by-product we obtain: a) new reduction conditions for the DS-I model
related to the
Ishimori-I (IM-I) model, b) the 2+1 dimensional zero curvature formulation
for the DS
equation.
\par
\bigskip\medskip
\noindent {\bf 1 Zero Curvature Condition} \bigskip
\par
\medskip
The crucial moment of our construction is a zero curvature condition in 2+1
dimensional
space
\par
$$
F_{\mu \nu }= \partial _{\mu }J_{\nu }-
\partial _{\nu }J_{\mu } + [J_{\mu }, J_{\nu }] = 0 , \eqno(1.1)$$
\noindent for $u(2) (u(1,1))$ Lie algebra valued connection $J_{\mu } ,
(\mu = 0,1,2)$.
Decomposing the algebra for the diagonal and off-diagonal parts \par
$$
J_{\mu } = J^{(0)}_{\mu }+ J^{(1)}_{\mu } ,\eqno(1.2)$$ and parametrizing
$$
J^{(0)}_{\mu }= {i\over4}(IW_{\mu } + \sigma _{3}V_{\mu }),\eqno(1.3a) $$
$$
J^{(0)}_{\mu }= \left(\matrix{ 0&-\kappa ^{2}\bar{q}_{\mu }\cr
q_{\mu}&0\cr}\right),
\eqno(1.3b)
$$
in terms of two $U(1)$ gauge potentials $W_{\mu }$ and $ V_{\mu }$, and the
$q_{\mu }$
matter fields,
we have the set of $U(1) \times U(1)$ gauge invariant equations $$
D_{\mu }q_{\nu }= D_{\nu }q_{\mu } ,\eqno(1.4a) $$
$$
[D_{\mu }, D_{\nu }] = - {i\over2}(\partial _{\mu }V_{\nu }- \partial _{\nu
}V_{\mu }) = -
2\kappa ^{2}(\bar{q}_{\mu }q_{\nu } - \bar{q}_{\nu }q_{\mu }) ,\eqno(1.4b)
$$
$$
\partial _{\mu }W_{\nu }- \partial _{\nu }W_{\mu }= 0 ,\eqno (1.4c) $$
Here the covariant derivative is defined as $D_{\mu } = \partial _{\mu } -
(i/2) V_{\mu }$.
\par
For the spatial part of the matter fields we introduce \par
$$
\psi _{\pm } = (2\pi ^{1/2})^{-1}(q_{1} \pm iq_{2}), \eqno(1.5)$$
and denote $D_{\pm }= D_{1}\pm iD_{2}$. Hence we have the system \par
$$
D_{0}\psi _{\pm } = (2\pi ^{1/2})^{-1}D_{\pm }q_{0} ,\eqno(1.6a) $$
$$
[D_{+}, D_{-}] = 8\pi \kappa ^{2}(|\psi _{+}|^{2} - |\psi _{-}|^{2}) ,
\eqno(1.6b)
$$
$$
[D_{0}, D_{\pm }] = -4\pi ^{1/2}\kappa ^{2}(\bar{q}_{0}\psi _{\pm } -
\bar{\psi }_{\pm }q_{0})
,\eqno(1.6c)
$$
$$
D_{+}\psi _{-}= D_{-}\psi _{+} ,\eqno(1.6d) $$
$$
\partial _{\mu }W_{\nu }- \partial _{\nu }W_{\mu }= 0 .\eqno(1.6e) $$
\noindent These equations can be derived from the Lagrangian density \par
$$
{\it L} = (32\pi \kappa ^{2})^{-1}\epsilon ^{\mu \nu \lambda }V_{\mu }
\partial _{\nu
}V_{\lambda } + \nu ^{-1}\epsilon ^{\mu \nu \lambda }W_{\mu }\partial _{\nu
}W_{\lambda }
$$
$$
+ i/2(\bar{\psi }_{+}D_{0}\psi _{+} - \psi _{+}\bar{D}_{0}\bar{\psi }_{+})
- i/2(\bar{\psi }_{-
}D_{0}\psi _{-} - \psi _{-}\bar{D}_{0}\bar{\psi }_{-}) $$
$$
+ i/2 \pi ^{-1/2}q_{0}\bar{(D_{-}}\bar{\psi_{+}} - \bar{D_{+}}\bar{\psi_{-}})
- i/2 \pi ^{-1/2} \bar{q_{0}}(D_{-}\psi _{+} - D_{+}\psi _{-}) ,
\eqno(1.7).$$
with the statistical parameters $\kappa ^{2}$ and $\nu$. It describes the
non-Abelian $
U(2)$ CS topological field theory,
in terms of parametrization (1.3-5) [11]. We note that at this stage the
gauge field $W$ is
decoupled
from the interaction with $\psi$ and $V$ fields. But, as we will see
below the mapping of the model (2.1) to zero curvature equations connects these
fields. Finally, let us emphasize that if for the Heisenberg model it was
sufficient to have $SU(2)$ zero curvature mapping [12], for the model
considered here
an additional gauge field, related to the velocity field ${\bf v}$ is
necessary.
\par
\bigskip \medskip
\noindent {\bf 2 Topological magnet in Tangent Space} \bigskip
\par
\medskip
To resolve the momentum problem in ferromagnets, the simple model of
de-localized
electrons was considered [9]. In this paper additional hydro-dynamical
variables which
describe the fermionic background were
introduced for the restoration of the correct linear momentum density. They
are the
density $n$ and the normal velocity ${\bf v}$ of the fermionic liquid. The
evolution of the
magnetization vector is described by the Landau- Lifshitz equation in a
moving frame
[10]. We observed that the vorticity of the flow
${\bf v}$ is proportional to the density of the topological charge for the
magnetic
configuration. This allowed us to formulate a simple model of ferromagnetic
continuum
with non-trivial background [8] \par
$$
\partial _{t}{\bf S} + v^{j}\partial _{j}{\bf S} = {\bf S \times \partial
}E/\partial {\bf S}
,\eqno(2.1a) $$
\noindent $$\partial _{i}v_{j}- \partial _{j}v_{i}= \theta {\bf S}(\partial
_{i}{\bf S \times \partial
}_{j}{\bf S}) , (i,j = 1,2),\eqno(2.1b)$$
where $E = E({\bf S},\partial {\bf S})$ - is an effective energy of the
Landau-Lifshitz
theory, ${\bf S}^{2}= 1$. For pure exchange interaction it has a simple
form \par
$$
E = (\partial _{1}{\bf S})^{2} + \alpha ^{2}(\partial _{2}{\bf S})^{2} .
\eqno(2.2)$$
\noindent The scalar product is defined by a diagonal metric tensor $g_{ij}
:A^{j}B_{j} =
g^{ij}A_{i}B_{j}$ with coefficients of exchange interaction constants (2.2)
\par
$$
g_{ij}=\hbox{ diag}(1,\alpha ^{2}) ,
\eqno(2.3)$$
and $\alpha ^{2} = \pm 1$.
Note that the parameter $\theta$ is defined by a microscopic model. But, as
was shown
in [8], only if $\theta = 1$ the model admits the Hirota bilinearization.
In this case as we
will see below
both an integrable reduction of the model and the statistical gauge field
compensation
exist. This is why we restrict the values only in this case.
\par If we consider the pair of $2\times 2$ matrix operators $$
L_{1} = \alpha I\partial_{2} + S\partial_{1} , $$
$$
L_{2} = I\partial_{t} + 2iS \partial_{1} ^{2} + (i\partial_{1}S +
i\alpha\partial_{2}S S - \alpha
S v_{2} + Iv_{1}) \partial_{1} , $$
with $S = \Sigma_{\alpha=1}^{3}S_{\alpha}\sigma_{\alpha} $, then from a
commutativity
condition
$$
[L_{1}, L_{2}] = 0 ,
$$
the system of equations follows
$$
\partial_{t}S + v^{j}\partial_{j}S = {1\over {2i}}[S, \partial^{j}
\partial_{j}S] +
(\partial^{1}v_{1} - \partial^{2}v_{2})S , \eqno(2.4a) $$
$$
\partial_{1}v_{2} - \partial_{2}v_{1} = -itr S[\partial_{1}S, \partial_{2}S] .
\eqno(2.4b)$$
In the case
of a restricted flow,
when the velocity field $\bf v$ satisfies the additional, ``incompressibility''
condition \par
$$
\partial _{1}v_{1} + \alpha ^{2}\partial _{2}v_{2} = 0 , \eqno(2.5)$$
the last term in eq.(2.4a) vanishes and we distinguish the model (2.1).
When eq.(2.5) is
resolved in terms of the stream function $\phi $ only,
\par
$$
v_{1} = \partial _{2}\phi , v_{2} = \alpha ^{2}\partial _{1}\phi , $$
the system (2.1) is reduced to the Ishimori model [7].
\par
In order to formulate the model in terms of tangent space variables we
diagonalize the
spin matrix
\par
$$
S = g \sigma _{3}g^{-1}.
\eqno(2.6)$$
Let us parametrize $g$ in such a way to extract $U(1) \times U(1)$ gauge
factors
(determined by $\beta$ and $\gamma$), leaving $S$ invariant
\par
$$
g = \left(\matrix{a&\bar{b}\cr {-\kappa}^{2}{b}&a\cr}\right)e^{i\sigma _{3}
\beta (x,t)+
i\gamma (x,t)} ,
\eqno(2.7)$$
where $a^{2}+ \kappa ^{2}|b|^{2} = 1 $, which defines the left current
\par
$$
J_{\mu } = g^{-1}\partial _{\mu }g,
\eqno(2.8)$$
being viewed as a connection, carrying the zero-curvature condition (1.1).
Equations of
motion (2.1) in terms
of this current have the form of
\par
$$
J^{(1)}_{0}= - v^{j}J^{(1)}_{j} + i(\partial ^{j}J^{(1)}_{j} + [ J^{(0)j},
J^{(1)}_{j}])\sigma _{3}
,\eqno(2.9a) $$
$$
\partial _{i}v_{j} - \partial _{j}v_{i}
= 2i tr([ J^{(1)}_{i}, J^{(1)}_{j}]\sigma _{3}) ,\eqno(2.9b) $$
and define the mapping of our model to zero curvature conditions (1.1). We
note that
eqs.(2.9) can be considered as a constraint on components of the current
$J$ and are
explicitly $U(1)$ gauge invariant. The dynamics of the system is now
described by the
zero curvature
equation (1.1). This peculiar behaviour is due to the first order time
derivatives in the
equations of motion (2.1). \par
The parametrization (1.2),(1.3) for $J$ has an explicit form in terms of
the spin variables
\par
$$
q_{\mu } = {\kappa \over{2}} S_{+}e^{2i\beta } \partial _{\mu }\ln
{S_{+}\over{1 + S_{3}}}
,\eqno(2.10a) $$
$$
r_{\mu } = -{1\over{2}} S_{-}e^{-2i\beta } \partial _{\mu }\ln
{S_{-}\over{1 + S_{3}}} = -
\kappa ^{2}\bar{q}_{\mu } ,\eqno(2.10b) $$
$$
V_{\mu } = i\kappa ^{2}{(S_{+}\partial _{\mu }S_{-} - S_{-}\partial _{\mu
}S_{+})\over{(1 +
S_{3})}} + 4\partial _{\mu }\beta ,\eqno(2.10c)
$$
$$
W_{\mu } = 4\partial _{\mu }\gamma .\eqno(2.10d) $$
\noindent For eqs.(2.9) we have
\par
$$
q_{0}= i(D^{j} + iv ^{j})q_{j} ,\eqno(2.11a) $$
$$
\partial _{i}v_{j} - \partial _{j}v_{i}
= -4i\kappa ^{2}(\bar{q}_{i}q_{j} - \bar{q}_{j} q_{i}) .\eqno(2.11b) $$
According to eq.(2.11b) one can express the transfer part of ${\bf v}$ as a
nonlocal
expression in terms of $q_{j}$. Then, by substituting ${\bf v}$
in eq.(2.11a), a complicated equation
for $q_{j}$ and $V_{j}$
arises. But,
as can be seen in eqs.(2.10), one cannot derive the longitudinal part of
${\bf v}$ by using
the local
gauge transformations related with parameter $\beta$. \par
{}From another point of view, we can connect an arbitrary function
$\gamma $ and the
gauge potential $W_{j}$ with $v_{j}$. The particular choice of this
connection allows
that eq.$(1.6e)$
will have the form of the ``incompressibility'' condition (2.5). If we put \par
$$
\phi = {1\over 4G_{0}}\gamma
,$$
where $G_{0}$ is an arbitrary constant, we find \par
$$
W_{1}= {\alpha ^{2}\over G_{0}} v_{2} , W_{2}= {1\over G_{0}} v_{1}, $$
and eq.(2.4) follows from eq.$(1.6e)$. The last relation between $\bf v$
and $\bf W$ is
just a duality transformation in two dimensions. \bigskip \medskip
\noindent
{\bf 3} $\alpha^{2} = -1$ {\bf Case}
\medskip \bigskip
\par
We consider the case $\alpha ^{2} = -1,$ with (+,-) signature for the
matrix $g _{ij}$ (2.3).
Eqs.(2.11) written in
terms of
$\psi _{\pm }$ given by (1.5), allows one to exclude $q_{0}$ from the
system (1.6). As a
result we obtain the set of equations $$
2iD_{0}\psi _{\pm } + (D^{2}_{+}+ D^{2}_{-})\psi _{\pm } + i(v_{+}D_{+}+
v_{-}D_{-})\psi
_{\pm }
{\pm} i\partial _{\pm }v_{\pm }\psi _{\pm }= 0 , \eqno(3.1a)$$
$$
\partial _{+}V_{-} - \partial _{-}V_{+} = 16\pi \kappa ^{2}i(|\psi
_{+}|^{2} - |\psi _{-}|^{2})
,\eqno(3.1b) $$
$$\partial _{0}V_{\pm } - \partial _{\pm }V_{0} = -8\pi \kappa ^{2}\{\psi
_{\pm }[(\bar{D}_{+}
- i v_{-})\bar{\psi }_{+}
+ (\bar{D}_{-} - i
v_{+})\bar{\psi }_{-}] +
$$
$$
\bar{\psi}_{\mp }[(D_{+} + iv_{+})\psi _{+} + (D_{-} + iv_{-}) \psi _{-}]\}
,\eqno(3.1c)$$
$$
D_{+}\psi _{-}= D_{-}\psi _{+} ,\eqno(3.1d) $$
$$\partial _{+}v_{-} - \partial _{-}v_{+} = 16\pi \kappa
^{2}i{\theta}(|\psi _{+}|^{2} - |\psi _{-
}|^{2}) , \eqno(3.1e)$$
$$\partial _{+}v_{-} + \partial _{-}v_{+} = 0 .\eqno(3.1f)$$ \par
It turns out that our model (2.1) can be described as a non-relativistic
field theory for a
pair of $\pm$
charged matter fields, interacting with the Abelian Chern-Simons gauge
field $V_{\mu}$
and the velocity field ${\bf v}$.
We note that, only due to the constraint (3.1f), equations (3.1a) for $\psi
_{\pm}$ formally
decouple. The matter fields are related only through the gauge field
$V_{\mu}$ and
velocity ${\bf v}$. \par
Equations (3.1b) and (3.e) are the Gauss laws of the Chern-Simons dynamics.
They
show that ${\bf V}$ and ${\bf v}$ are determined by the matter density. In
this theory the
excitations carrying charge $$
Q = \int (|\psi _{+}|^{2} - |\psi _{-}|^{2}) d^{2}x , $$
also possess magnetic and hydrodynamical fluxes, determined by statistical
parameters
$\kappa^{2}$ and $\theta \kappa^{2}$ respectively. When $\theta = 1$ , the
fluxes
coincide and can compensate one another. In this case
we can introduce
new, irrotational, gauge field
\par
$$
{\cal A} = {\bf V} - {\bf v} .
\eqno(3.2)$$
Note that
it transforms correctly as a $ U(1)$
gauge field under the shift of $\beta$ in eq.(2.10). The related covariant
derivative will be
denoted as \par
$$
{\cal D} = \partial - i/2 {\cal A} .
\eqno(3.3)$$
Then the system (3.1) becomes
\par
$$
2iD_{0}\psi _{\pm } + ({\cal D}^{2}_{+}+ {\cal D}^{2}_{-}) \psi _{\pm }+
[{1\over{4}}
(v^{2}_{+} + v^{2}_{-} ) \pm {i\over{2}} (\partial _{+}v_{+}- \partial
_{-}v_{-})] \psi _{\pm }= 0 ,
\eqno(3.4a)$$
$$
\partial _{+}{\cal A}_{-} - \partial _{-}{\cal A}_{+} = 0 ,\eqno(3.4b) $$
$$\partial _{0}{\cal A}_{\pm } - \partial _{\pm }V_{0} = - \dot{v}_{\pm }
-8\pi \kappa ^{2}\{\psi
_{\pm }[(\bar{\cal D}_{+} - {i\over{2}}v_{-})\bar{\psi }_{+}
+ (\bar{\cal D}_{-} - {i\over{2}}
v_{+})\bar{\psi }_{-}] +
$$
$$
\bar{\psi}_{\mp }[({\cal D}_{+} + {i\over{2}} v_{+})\psi _{+} + ({\cal
D}_{-} + {i\over{2}} v_{-
})\psi _{-}]\} ,\eqno(3.4c)$$ $$
({\cal D}_{+} - {i\over{2}} v_{+})\psi _{-} = ({\cal D}_{-} - {i\over{2}}
v_{-})\psi _{+}
,\eqno(3.4d)$$ $$\partial _{+}v_{-} - \partial _{-}v_{+} = 16\pi \kappa
^{2}i(|\psi _{+}|^{2} -
|\psi _{-}|^{2}) ,\eqno(3.4e)$$ $$
\partial _{+}v_{-} + \partial _{-}v_{+} = 0 .\eqno(3.4f) $$
{}From eq.(3.4a) the charge +,- conservation laws follow
\par
$$
\partial _{0}|\psi _{\pm }|^{2} + \partial _{+}J^{(\pm )}_{+} + \partial
_{-}J^{(\pm )}_{-} = 0
,\eqno(3.5)$$
where the currents are
\par
$$
{\bf J}^{(\pm )} = i/2(\psi _{\pm }\bar{{\cal D}}\bar {\psi}_{\pm } -
\bar{\psi}_{\pm } {\cal D}
\psi _{\pm } ). \eqno(3.6)$$
But due to the constraint (3.4d)
one of the global $U(1)$
symmetries is broken and we have only $U(1)$transformations \par
$$
\psi _{+} -> \psi _{+}e^{i\alpha} , \psi _{-} -> \psi _{-}e^{i\alpha} .
\eqno(3.7)$$
Moreover, the system (3.4) is invariant under $U(1)$ local gauge transformation
\par
$$
\psi _{\pm } -> \psi _{\pm }e^{i\alpha}
,\eqno(3.8)$$
$$
{\cal A}_{\pm } -> {\cal A}_{\pm }
+ 2 \partial _{\pm }\alpha , V_{0} -> V_{0} + 2 \partial _{0}\alpha
,$$
connected with an arbitrariness of parameter $\beta $ in (2.10).
According
to (3.4b), the planar gauge field components ${\cal A} _{\pm}$ have a
vanishing field
strength, and must be of a pure gauge form. We
can solve this in terms of real function $\lambda $, \par
$$
{\cal A}_{j}= \partial _{j}\lambda
, (j = 1,2) ,\eqno(3.9)$$
and introduce the gauge-invariant matter fields \par
$$
\Psi _{\pm }= \psi _{\pm }e^{-{i\over{2}}\lambda (x_{1},x_{2},t)}
.\eqno(3.10)$$
\noindent Hence the system (3.4) simplifies $$2i(\partial _{0} -
{i\over{2}} {\cal V}_{0})\Psi
_{\pm } + (\partial ^{2}_{+}+ \partial ^{2}_{-})\Psi _{\pm } + [{1\over{4}}
(v^{2}_{+} + v^{2}_{-}
)
\pm {i\over{2}} (\partial _{+}v_{+}- \partial _{-}v_{-})] \Psi _{\pm }= 0
,\eqno(3.11a)$$
$$\dot{v}_{\pm } = \partial _{\pm }{\cal V}_{0} - 8\pi \kappa ^{2}\{\Psi
_{\pm }(\partial _{-
}\bar{\Psi }_{+} + \partial _{+}\bar{\Psi }_{-}) + \bar{\Psi }_{\mp
}(\partial _{+}\Psi _{+} +
\partial _{-}\Psi _{-})\}
\pm
$$
$$
4\pi \kappa ^{2}iv_{\mp } (|\Psi _{+}|^{2}- |\Psi _{-}|^{2}) ,\eqno(3.11b)$$
$$(\partial _{+} - {i\over{2}} v_{+})\Psi _{-} = (\partial _{-} -
{i\over{2}} v_{-})\Psi _{+}
,\eqno(3.11c)$$ $$\partial _{+}v_{-} - \partial _{-}v_{+} = 16\pi \kappa
^{2}i(|\Psi _{+}|^{2} -
|\Psi _{-}|^{2}) ,\eqno(3.11d)$$ $$
\partial _{+}v_{-} + \partial _{-}v_{+} = 0 .\eqno(3.11e) $$
Note that the statistical gauge field $\bf V$ has completely disappeared
from our system.
Only the Gauss law for hydrodynamical vorticity (3.11d) remains. \par
After introducing new fields
$$
{\cal A}^{(\pm )}_{0} = {\cal V}_{0}+ {1\over{4}} (v^{2}_{+} + v^{2}_{-} )
\pm {i\over{2}}
(\partial _{+}v_{+}- \partial _{-}v_{-}), \eqno(3.12)$$
and performing long but straightforward
calculations, we
surprisingly obtain that eqs. (3.11a) and (3.11b) become Davey-Stewartson
equations
\par
$$
2i\partial _{0}\Psi _{\pm } + (\partial ^{2}_{+}+ \partial ^{2}_{-})\Psi _{\pm
}+
{\cal A}^{(\pm )}_{0} \Psi _{\pm }= 0 ,\eqno(3.13a) $$
$$
\partial _{+}\partial _{-}{\cal A}^{(\pm )}_{0} = 8\pi \kappa ^{2}(\partial
^{2}_{+}
+ \partial ^{2}_{-})|\Psi _{\pm }|^{2} ,\eqno(3.13b) $$
\noindent for the pairs
$(\Psi _{+},{\cal A}^{(+)}_{0}),(\Psi _{-},{\cal A}^{(-)}_{0})$. These
equations are known
as the DS-II equations and are an integrable system [13]. The potentials
${\cal A}^{(\pm
)}_{0}$ are connected by the relation \par
$$
{\cal A}^{(+)}_{0}- {\cal A}^{(-)}_{0}
= i(\partial _{+}v_{+}- \partial _{-}v_{-}) , \eqno(3.14)$$
\noindent and the functions $\Psi _{\pm }$ by $(3.11c-e)$ constraints.
These constraints
define ${\bf v}$ as a nonlocal form of $\Psi _{+}$ and $\Psi _{-}$ ,
producing a
complicated reduction between these functions. \par
The zero-curvature potentials (1.2),(1.3), related to the system (3.11),
can be obtained by
the local gauge transformation $h = \exp (-i/4 \lambda \sigma _{3})$
and have the form
\par
$$
J^{(0)}_{0}= {i\over{4}}(IW_{0} + \sigma _{3}{\cal V}_{0}), J^{(0)}_{\pm }
= {i\over{4}}(\pm
I{i\over{G_{0}}}v_{\pm } + \sigma _{3}v_{\pm }), \eqno(3.15)$$
$$J^{(1)}_{\pm }= 2\pi ^{1/2} \left(\matrix{0&-\kappa ^{2} \bar{\Psi }_{\mp
}\cr
\Psi_{\pm}&0\cr}\right ), J^{(1)}_{0} = \left(\matrix{ 0&-\kappa
^{2}\bar{s}_{0}\cr
s_{0}&0\cr}\right ),$$ $$
s_{0} = i\pi ^{1/2}\{(\partial _{+}+ {i\over{2}} v_{+})\Psi _{+} +
(\partial _{-}+ {i\over{2}} v_{-
})\Psi _{-}\}. $$
\bigskip \medskip
\noindent {\bf 4 Self-Dual Chern-Simons Solitons} \par
\medskip
\bigskip
Let us consider a particular but physically important subclass of solutions
of the system
$(3.11c-e)$ with the evolution (3.13). If one of the functions $\Psi _{\pm
}$ vanishes
the (anti-)holomorphicity
condition for the original spin variables (2.10) is satisfied. Let
\par
$$
\Psi _{-} = 0 .
\eqno(4.1)$$
\noindent Then as follows
$\partial _{-}\zeta = 0,$ where $\zeta $ is the stereographic projection of the
spin vector {\bf S}. In this case the system $(3.11c-d)$ reduces to the
self-dual Chern-Simons system
\par
\noindent $$(\partial _{-} - {i\over{2}} v_{-})\Psi _{+} = 0 , \eqno(4.2a)$$
$$\partial _{+}v_{-} - \partial _{-}v_{+} = 16\pi \kappa ^{2}i |\Psi
_{+}|^{2} .\eqno(4.2b)$$
We can call it {\it hydrodynamical}, since the velocity of the planar flow
plays
the role of the  statistical gauge field. This system is connected with
the Liouville equation [1].
If we represent the velocity
\par
$$
v_{1}= \partial _{2}\phi + \partial _{1}\chi , v_{2} = -\partial _{1}\phi +
\partial _{2}\chi
,\eqno(4.3)$$
\noindent in terms of two real functions for the flow, the {\it stream
function}
$\phi $ and the {\it velocity potential}
$\chi $, then
from (4.2a) the
general form of $\Psi _{+}$ follows
\par
$$
\Psi _{+}= \exp (-{1\over{2}} \phi + {i\over{2}} \chi )F(\bar{\eta }) =
\rho_{+} e^{{i\over
{2}}\omega}
,\eqno(4.4)$$
\noindent where $F(\bar{\eta })$ is an arbitrary anti-holomorphic function,
$\eta = x_{1} +
ix_{2}$.
Substituting the $\Psi _{+}$ in eq.(4.2b), we find for $\rho_{+} = |\Psi
_{+}|^{2}$
the Liouville equation:
\par
$$
(\partial ^{2}_{1} + \partial ^{2}_{2})\ln \rho_{+} = - 8\pi \kappa
^{2}\rho_{+} .
\eqno(4.5)$$
\par This equation is conformal invariant \par
$$
\eta = f(\tilde{\eta }), \bar{\eta } = \bar{f}(\bar{\tilde{\eta}}) , $$
$$\ln\tilde{\rho }(\tilde{\eta },\bar{\tilde{\eta}}) = \ln\rho [f(\tilde{\eta
}),\bar{f}(\bar{\tilde{\eta}})] + \ln (f'\bar{f}') ,\eqno(4.6) $$
and has regular, nonnegative solutions
only for the compact
spin model when $\kappa ^{2} = 1.$ The general solution is \par
$$
\rho_{+} = |\Psi _{+}|^{2}
= {1\over \pi}{{|\partial _{\bar\eta }\zeta |^{2}}\over{(1 + \kappa
^{2}|\zeta |^{2})
^{2}}}
.\eqno(4.7)$$
\par According to eq.$(3.11e)$ the function $\chi $ should be harmonic
\par
$$
\Delta \chi = 0,
\eqno(4.8)$$
\noindent as well as the phase
of function $\Psi _{+}$. The total
phase of function $\Psi_{+}$ should satisfy the regularity requirements for
velocity ${\bf
v}$ and the single-valuedness [1]. \par
The $N$ - vortex-soliton solution
\par
$$
\zeta (\bar{\eta} ) = \Sigma_{n=1}^{N}{c_{n}\over{\bar{\eta} - \bar{\eta}_{n}}}
,
\eqno(4.9)$$
is defined by $4n$ real parameters,
coded in the complex
$c_{n}(t)$
and $\bar{\eta}_{n}(t)$, describing the scale, phase and position of
solitons on the plane.
The corresponding topological charge is $Q = + N$. In the case $\Psi_{+} =
0$, we have
$\partial_{+}\zeta = 0$
and the same Liouville equation (4.5)
for $\rho_{-}$. Replacing $\bar{\eta}$
to $\eta$ in the solution (4.9) we can obtain the $N$ - antivortex soliton
having a charge $Q = - N$.
The evolution of these solitons should follow to the DS equations (3.13).
It provides the
time dependence for parameters $c_{n}(t)$
and $\bar{\eta}_{n}(t)$. Combining properly these parameters one can obtain
integrals of motion.
Further details of integrable soliton dynamics will be published separately.
Here we note only that
in paper [14] the reduction of the DS equation to the Liouville equation
was
considered along with colliding closed curves of singularities. In our
model these results
are applicable to noncompact spin algebra when $\kappa^{2} = -1$. Moreover, the
self-dual Chern-Simons system (4.2) is related to a more general reduction for
the DS  equations than the Liouville equation.
\bigskip \medskip \noindent
{\bf 5} $\alpha ^{2} = 1$ {\bf Case}
\par \bigskip \medskip
In this section we briefly describe the $\alpha ^{2} = 1$ reduction of the
system (2.1). In this case instead of (1.5) we introduce new
complex fields
\par
$$
\chi _{\pm }= 1/2\pi ^{-1/2}(q_{1} \pm q_{2}), \bar{\chi }_{\pm }= 1/2\pi
^{-1/2}(\bar{q}_{1}
\pm \bar{q}_{2}) ,\eqno(5.1)$$
and define the covariant derivative and velocity fields as
\par
$$
D_{\pm } = D_{1} \pm D_{2} , v_{\pm } = v_{1} \pm v_{2} .$$
\noindent Then instead of eqs.(1.6) from the zero curvature equations we have
\par
$$
D_{0}\chi _{\pm } = (2\pi ^{1/2})^{-1}D_{\pm }q_{0} ,\eqno(5.2a) $$
$$
[D_{+}, D_{-}] = 8\pi \kappa ^{2}(\bar{\chi }_{-}\chi _{+} - \bar{\chi
}_{+}\chi _{-})
,\eqno(5.2b) $$
$$
[D_{0}, D_{\pm }] = -4\pi ^{1/2}\kappa ^{2}(\bar{q}_{0}\chi _{\pm } -
\bar{\chi }_{\pm }q_{0})
,\eqno(5.2c)
$$
$$D_{+}\chi _{-}= D_{-}\chi _{+} ,\eqno(5.2d)$$ and for the equations of
motion in tangent
space (2.9) \par
$$
q_{0} = i\pi ^{1/2}\{(D_{+} + iv_{+})\chi _{+} + (D_{-} + iv_{-})\chi
_{-}\} ,\eqno(5.3a) $$
$$
\partial _{+}v_{-} - \partial _{-}v_{+} = 16i\pi \kappa ^{2}(\bar{\chi
}_{-}\chi _{+} - \bar{\chi
}_{+}\chi _{-}) .\eqno(5.3b) $$
Again substituting $q_{0}$ in system (5.2) and introducing the new gauge
field as in
eq.(3.2) we find that it must have zero strength, ${\cal A}_{j}= \partial
_{j}\lambda $. \par
In terms of gauge invariant variables
$$
X_{\pm }= \chi _{\pm }e^{-{i\over{2}} \lambda} \eqno(5.4)$$
\noindent the system becomes
\par
$$2i(\partial _{0}- {i\over{2}} {\cal V}_{0})X_{\pm } + (\partial ^{2}_{+}+
\partial ^{2}_{-
})X_{\pm } + [{1\over{4}} (v^{2}_{+} + v^{2}_{-} )
\pm {i\over{2}} (\partial _{+}v_{+}- \partial _{-}v_{-})] X_{\pm }= 0
,\eqno(5.5a)$$
$$
\dot{v}_{1}= \partial _{1}{\cal V}_{0}
- 4\pi \kappa ^{2}\{(\bar{X}_{+}+
\bar{X}_{-})[(\partial _{+}+ {i\over{2}} v_{+})X_{+} + (\partial _{-}+
{i\over{2}} v_{-})X_{-}] +
c.c \}, ,\eqno(5.5b)$$
$$
\dot{v}_{2}= \partial _{2}{\cal V}_{0} - 4\pi \kappa ^{2}\{(\bar{X}_{+}-
\bar{X}_{-})[(\partial _{+}+ {i\over{2}} v_{+})X_{+} + (\partial _{-}+
{i\over{2}} v_{-})X_{-}] +
c.c \} ,\eqno(5.5c)$$
$$(\partial _{+} - {i\over{2}} v_{+})X_{-} = (\partial _{-} - {i\over{2}}
v_{-})X_{+}
,\eqno(5.5d)$$ $$
\partial _{+}v_{-} - \partial _{-}v_{+} = 16i\pi \kappa
^{2}(\bar{X}_{-}X_{+} - \bar{X}_{+}X_{-})
,\eqno(5.5e)$$ $$
\partial _{+}v_{-} + \partial _{-}v_{+} = 0 .\eqno(5.5f)$$ \par
In contrast to the previous case, with $\alpha^{2} = -1$, the potentials $$
{\cal A}^{(\pm )}_{0} = {\cal V}_{0}+ {1\over{4}} (v^{2}_{+} + v^{2}_{-} )
\pm {i\over{2}}
(\partial _{+}v_{+}- \partial _{-}v_{-}) $$
are no longer real functions.
Then we obtain the DS - I equations
$$
2i\partial _{0}X_{\pm } + (\partial ^{2}_{+}+ \partial ^{2}_{-})X_{\pm }+
{\cal A}^{(\pm )}_{0}
X_{\pm }= 0 ,\eqno(5.6a) $$
$$
-2i\partial _{0}\bar{X}_{\mp } + (\partial ^{2}_{+}+ \partial ^{2}_{-})
\bar{X}_{\mp }+ {\cal
A}^{(\pm )}_{0} \bar{X}_{\mp }= 0 ,\eqno(5.6b) $$
$$
\partial _{+}\partial _{-}{\cal A}^{(\pm )}_{0}= 8\pi \kappa ^{2}(\partial
^{2}_{+} +
\partial ^{2}_{-}) X_{\pm }\bar{X}_{\mp } .\eqno(5.6c) $$
\par
Comparing these equations with the well known ones in the literature we
introduce the
definition
\par
$$
Q = X_{+} , R = \bar{X}_{-} .
\eqno(5.7)$$
Thus we have the DS-I equations
$$
2i\partial _{0}Q + (\partial ^{2}_{+}+ \partial ^{2}_{-})Q + {\cal
A}^{(+)}_{0} Q = 0
,\eqno(5.8a)
$$
$$
-2i\partial _{0}R + (\partial ^{2}_{+}+
\partial ^{2}_{-})R + {\cal A}^{(+)}_{0} R = 0 ,\eqno(5.8b) $$
$$
\partial _{+}\partial _{-}{\cal A}^{(+)}_{0}= 8\pi \kappa ^{2}(\partial
^{2}_{+} + \partial ^{2}_{-
}) RQ ,\eqno(5.8c) $$
supplied with constraints
$$
(\partial _{+} - {i\over{2}} v_{+})\bar{R} = (\partial _{-} - {i\over{2}}
v_{-})Q ,\eqno(5.9a)$$
$$\partial _{+}v_{-} - \partial _{-}v_{+} = 16i\pi \kappa ^{2}(RQ -
\bar{R}\bar{Q})
.\eqno(5.9b)$$ \par
{}From the above equations we can conclude that IM-I
model is connected with DS-I
equation (5.8) but with a new reduction condition (5.9), instead of the
usual one
$$
\bar{R} = \pm Q .
\eqno(5.10)$$
\par If we suppose that (5.10) is
satisfied in addition to (5.9), then ${\bf v} = 0$. Recovering ${\cal
A}^{(+)}_{0}$ from
eq.(5.8c) we find the one-dimensional
Nonlinear Schr\"odinger equation
\par
$$
i{Q} + \partial ^{2}_{1}Q
+ 8\pi \kappa ^{2}(|Q|^{2}- \rho _{1}(t)x_{1}- \rho _{0}(t))Q = 0. $$
\bigskip
\medskip
\noindent
{\bf 6 Conclusion}
\bigskip
\medskip
\par
In conclusion I like to emphasize some points. First, as mentioned above,
the original
model for $\theta = 1$ is an integrable. It admits the linear
problem as a commutativity condition between two operators $L_{1}$ and $L_{2}$.
The linear problem allows one to apply the whole machinery of the
$\bar\partial$
- problem to construct several exact solutions [18]. These solutions could
have direct
physical applications since DS equations arise in hydrodynamical wave
phenomena. It
is important to note the appearance in this context of the self-dual
Chern-Simons system
(4.2), usually associated with the anyons and exotic statistics. \par
The linear
problem of IM can be gauge related with the one of the DS equation [15,16].
And the last
linear problem can be reduced from the self-dual Yang-Mills system with an
infinite
dimensional gauge group [17]. On the other hand it has been shown in the
present paper
that the zero-curvature
formulation of the DS equations can be derived on the usual
finite-dimensional Lie
algebra. Moreover, these zero curvature conditions are the classical
equations of motion
for the 2+1 dimensional Chern-Simons Topological Quantum Field Theory
(TQFT). This
allows the mapping of the model to the TQFT with a
suitable gauge group. Based on the ideas of [19,20,21], we can expect some
relationship
with predictions of the TQFT and the properties of the Braid group
representations. We
hope that full 2+1- dimensional solutions for our model will provide a
description of
the moduli space beyond the perturbative Manton's approach for the slowly
moving
monopoles [19 ]. \par
Finally, I stress that the idea of vorticity compensation for the two gauge
fields may be
applied to other $\sigma$- models and it will be interesting to
analyze the anyonic behaviour for such systems. Generally, we can expect
that the local
rotation in the anyon fluid will produce
the statistical
transmutation connected with the vorticity of the flow. Therefore the
integrability
properties of the model can only appear for special
values of
the vorticity, compensating the anyonic statistical field. \bigskip
\medskip
\noindent
{\bf Acknowledgments}
\bigskip
\medskip
\par
The author would like to thank Professor Abdus Salam, the International Atomic
Energy Agency and UNESCO for hospitality at the International Centre for
Theoretical Physics, Trieste.
He would also like to thank Professor S. Randjbar-Daemi for his
invaluable support.
\vfill\eject

\noindent
{\bf References}
\medskip
\bigskip
\par \noindent
[1] R. Jackiw and S.Y.Pi, Phys. Rev. Lett. {\bf 64}, 2969 (1990); Phys.
Rev.{\bf
D42}, 3500  (1990).
\par \noindent
[2] F. Wilczek, {\it Fractional Statistics and Anyon Superconductivity} , World
Scientific,  Singapore, 1990.
\par \noindent
[3] {\it The Quantum Hall Effect}, edited by S. Girvin and R. Prange,
Springer-Verlag, New-York, 1990.
\par \noindent
[4] A.P. Balachandran et al., {\it Hubbard Model and Anyon Superconductivity},
Lecture  Notes in Physics, {\bf Vol.38}, World Scientific, Singapore, 1990.
\par \noindent
[5] L. Martina, O.K. Pashaev and G. Soliani, Mod. Phys. Lett. {\bf A8}, 3241
(1993).
\par \noindent
[6] R. Jackiw and S.Y. Pi, Progr. Theor. Phys. Suppl. {\bf 107}, 1 (1992). \par
\noindent [7] Y. Ishimori, Progr. Theor. Phys. {\bf 72}, 33 (1984). \par
\noindent [8] L. Martina, O.K. Pashaev and G. Soliani, J. Phys. A: Math. Gen.
{\bf 27}, 943 (1994).
\par \noindent
[9] G.E. Volovik, J. Phys. C: Solid State Phys. {\bf 20}, L83 (1987). \par
\noindent [10] L. Martina, O.K. Pashaev and G. Soliani, Theor. Math. Phys. {\bf
99}, 462 (1994).
\par \noindent
[11] L. Martina, O.K. Pashaev and G. Soliani, {\it Topological Field Theory on
Symmetric  Spaces and Planar Ferromagnets}, Lecce University preprint DF -
1/01/95.
\par \noindent
[12] L. Martina, O.K. Pashaev and G. Soliani, Phys. Rev. B {\bf 48}, 15 787
(1993).
\par \noindent
[13] A. Davey and K. Stewartson, Proc. R. Soc. London {\bf A338}, 101 (1974);
D. Anker and  N. Freeman, Proc. R. Soc. London {\bf A360}, 529 (1978). \par
\noindent [14] V.A. Arkadiev, A.K. Pogrebkov and M.C. Polivanov, Inverse
Problems
{\bf 5}, L1  (1989).
\par \noindent
[15] V.D. Lipovskii and A.V. Shirokov, Funk. Anal. Priloz. {\bf 23}, 65 (1989).
\par \noindent [16] R.A. Leo, L. Martina and G. Soliani, J. Math. Phys.
{\bf 33},
1515 (1992) \par \noindent [17] M.J. Ablowitz, S. Chakravarty and L.A.
Takhtajan, Comm. Math. Phys. {\bf 158}, 289  (1993).
\par \noindent
[18] B.G. Konopelchenko and B.T. Matkarimov, J. Math. Phys. {\bf 31}, 2737
(1990).
\par \noindent
[19] N. Manton, Phys. Lett. {\bf B110}, 54 (1982) \par \noindent
[20] L. Baulieu and B. Grossman, Phys. Lett. {\bf B 214}, 223 (1988). \par
\noindent [21] Y.S. Wu, Phys. Rev. Lett. {\bf 52}, 2103 (1984). \end